\documentclass[prl, twocolumn]{revtex4}%
\usepackage{amssymb}
\usepackage{amsmath}
\usepackage{amsfonts}
\usepackage{graphicx}%
\begin{document}
\title{Kinetics of a superconductor excited with a femtosecond optical pulse}
\author{V.V. Kabanov, J. Demsar, D. Mihailovic}
\affiliation{Department for Complex Matter, Jozef Stefan Institute, Jamova 39, SI-1000,
Ljubljana, Slovenia}
\date{\today}

\begin{abstract}
Superconducting state dynamics following excitation of a superconductor with a
femtosecond optical pulse is studied in terms of a phenomenological Rothwarf
and Taylor model. Analytical solutions for various limiting cases are
obtained. The model is found to account for the intensity and temperature
dependence of both photoinduced quasiparticle density, as well as
pair-breaking and superconducting state recovery dynamics in conventional as
well as cuprate superconductors.

\end{abstract}

\maketitle

In recent years numerous studies of non-equilibrium carrier dynamics in
superconductors (SC) have been performed utilizing femtosecond real-time
techniques [1-18]. Research focused on the identification of relaxation
processes and direct measurements of the relaxation times. One of the open
issues at the moment is whether cuprates are in the so called phonon
bottleneck regime as conventional SCs\cite{escape}, or in the weak bottleneck
regime, where relaxation is governed by the biparticle recombination kinetics
\cite{mercury,orenstein,Kaindl}. The theoretical model that has been most
commonly used to interpret the dynamics is a phenomenological Rothwarf-Taylor
(RT) model which describes the evolution of quasiparticle (QP) and high
frequency phonon (HFP) populations via a set of two non-linear differential
equations\cite{roth}, which were shown recently\cite{ovchkres} to follow from
the general set of kinetic equations for a SC\cite{eliash}. While the RT model
has been known for almost 40 years, no rigorous attempt to solve it has been
made thus far, and neither has a comparison to the experimental data been made.

In this Letter we present a detailed study of the evolution of the SC state
following excitation by ultrashort laser pulse using the RT model. We have
obtained analytical solutions of the model in the limit of a strong and a weak
bottleneck, which are in excellent agreement with numerical simulations. The
solutions enable comparison of the model to the experimental results. We show
that RT model can account for most of the experimental observations in
conventional as well as cuprate SC, and that both conventional and cuprates
SCs are in the strong bottleneck regime, where SC state recovery is governed
by the HFP decay dynamics.

Rothwarf and Taylor have pointed out that the phonon channel should be
considered when studying the SC relaxation\cite{roth}. When two QPs with
energies $\geq\Delta$, where $\Delta$ is the SC gap, recombine a HFP
($\omega>2\Delta$) is created. Since HFP can subsequently break a Cooper pair
creating two QPs the SC recovery is governed\ by the decay of the HFP
population. The dynamics of QP and HFP populations is determined
by\cite{roth}:
\begin{align}
dn/dt  &  =I_{0}+\eta N-Rn^{2}\nonumber\\
dN/dt  &  =J_{0}-\eta N/2+Rn^{2}/2-\gamma(N-N_{T}) \label{RTeq}%
\end{align}
Here $n$ and $N$ are concentrations of QPs and HFPs, respectively, $\eta$ is
the probability for pair-breaking by HFP absorption, and $R$ the bare QP
recombination rate with the creation of a HFP. $N_{T}$ is the concentration of
HFP in thermal equilibrium at temperature $T$, and $\gamma$ their decay rate.
$I_{0}$ and $J_{0}$ represent the external sources of QPs and HFPs,
respectively \cite{mgb2}. Physically $\gamma$ is governed\ by the fastest of
the two processes: anharmonic decay of HFP ($\omega<2\Delta$ phonons do not
have sufficient energy to break Cooper-pairs) \cite{kabanov} and the diffusion
of HFP into the substrate \cite{escape}. While the main T-dependence in
$\gamma$ appears near T$_{c}$\cite{kabanov}, some T-dependence is expected at
low-T as well. However, as shown in Refs.\cite{Maris,Mihailovic} this
T-dependence is very weak in the temperature range of interest and we can
consider $\gamma$ to be T-independent.

Since $\eta/R$ has the dimensionality of concentration we introduce
dimensionless QP and HFP concentrations, $q\equiv Rn/\eta$ and $p\equiv
RN/\eta$, while $\theta=\eta t$ and $\tilde{\gamma}\equiv\gamma/\eta$ are the
dimensionless time and HFP decay rate, respectively. Since a\ femtosecond
optical pulse is usually shorter than characteristic timescales of SC
dynamics, $I_{0}$ and $J_{0}$ could be approximated by $\delta$-functions
leading to the initial concentrations of QPs $q_{0}=Rn_{0}/\eta$ and HFP
$p_{0}=RN_{0}/\eta$ \cite{mgb2}. As a result Eqs. (\ref{RTeq}) reduce to
\begin{align}
dq/d\theta &  =p-q^{2}\label{RTdim1}\\
dp/d\theta &  =-p/2+q^{2}/2-\tilde{\gamma}(p-p_{T}). \label{RTdim2}%
\end{align}
with the initial conditions $p(0)=p_{0}$, $q(0)=q_{0}$. Here $p_{T}%
=RN_{T}/\eta$ is the dimensionless concentration of HFPs in thermal
equilibrium. Thermal equilibrium concentrations of HFPs and QPs ($q_{T}$)
satisfy the detailed balance equation $p_{T}=q_{T}^{2}$.

We investigated\ various limiting cases. We refer to the situation when the
photoinduced density is small $(p_{0}-p_{T}),(q_{0}-q_{T})\ll1$ as a weak
perturbation regime, while the opposite case is referred to as a strong
perturbation. In case $\tilde{\gamma}\ll1$ we have a strong bottleneck
(quasi-equilibrium between QPs and HFPs is established, and SC state recovery
is governed entirely by $\tilde{\gamma}$) while in the opposite case
($\tilde{\gamma}\gg1$) SC is in a weak bottleneck regime.

It is useful to estimate the values of thermal QP and HFP concentrations and
compare them with $\eta/R$. If $D(\omega)=9\nu\omega^{2}/\omega_{D}^{3}$ is
the phonon density of states with $\nu$ the number of atoms per unit cell and
$\omega_{D}$ the Debye energy, it follows that $N_{T}=\frac{36\nu\Delta^{2}%
T}{\omega_{D}^{3}}\exp(-2\Delta/k_{B}T)$. On the other hand the QP density per
unit cell at temperature T is given by \cite{kabanov}:%
\begin{equation}
n_{T}\simeq N(0)\sqrt{2\pi\Delta k_{B}T}\exp(-\Delta/T). \label{nTBCS}%
\end{equation}
Here $N(0)$ is the electronic density of states per unit cell. From $\eta/R= $
$n_{T}^{2}/N_{T}$ follows that $\eta/R=\frac{N(0)^{2}\pi\omega_{D}^{3}}%
{18\nu\Delta}$. Regardless of the value of $\eta/R$, it follows that the high
perturbation limit is reached when the photoexcitation density is close to the
density required for complete depletion of the SC state. To show this, we
estimate the temperature $\widetilde{T}$ where $q_{\tilde{T}}\approx1$
($n_{\tilde{T}}$ $\approx\eta/R$ ). It follows that $\widetilde{T}$
$\approx\Delta/\ln(\frac{\Delta^{2}\nu E_{F}}{\omega_{D}^{3}})$; thus $n_{T}$
is comparable to $\eta/R$ only in the close vicinity of $T_{c}$.

\textbf{Strong bottleneck regime. }Let us consider the solution of
Eqs.(\ref{RTdim1},\ref{RTdim2}) for the case when $\tilde{\gamma}\ll1$. In
this case two distinct regimes can be defined: (I) prebottleneck dynamics,
describing the processes on timescale much shorter than $\tilde{\gamma}^{-1}$,
and (II) the SC state recovery at $\theta\geq1/\tilde{\gamma}$.

\emph{(I) Prebottleneck dynamics }describes the short timescale evolution of
concentrations $p(\theta)$ and $q(\theta)$ preceding the relaxation. For such
short times we can neglect the last term in Eq.(\ref{RTdim2}). It leads to the
conservation law $q(\theta)+2p(\theta)=q_{0}+2p_{0}$ and exact solutions for
QP and HFP populations, respectively, in the pre-bottleneck regime are
obtained\cite{mgb2}:
\begin{align}
q(\theta)  &  =\left[  -\frac{1}{4}-\frac{\xi^{-1}}{2}+\frac{\xi^{-1}}%
{1-K\exp{(-\theta/\xi)}}\right] \label{PreQ}\\
p(\theta)  &  =\frac{1}{2}\left[  \frac{1}{8}+\frac{\xi^{-1}}{2}+\frac
{\xi^{-2}}{2}-\frac{\xi^{-1}}{1-K\exp{(-\theta/\xi)}}\right]  \label{PreP}%
\end{align}
Here $\xi^{-1}=\sqrt{1/4+4p_{0}+2q_{0}}$ and $K=\frac{(4q_{0}+1)-2\xi^{-1}%
}{4q_{0}+1)+2\xi^{-1}}$. At times $\theta>1$ concentrations $q(\theta)$ and
$p(\theta)$ reach the quasi stationary solution
\begin{equation}
q_{s}=\frac{1}{4}(\sqrt{1+16p_{0}+8q_{0}}-1)\text{ \ ; \ \ \ \ \ }p_{s}%
=q_{s}^{2}. \label{Stat}%
\end{equation}
The prebottleneck dynamics depends on the initial conditions, and has two
distinct regimes characterized by the parameter $K$. The regime $0<K\leq1$
corresponds to the situation when $q_{0}>q_{s}$, and $q(\theta)$ rapidly
decreases during the formation of the bottleneck - Fig. 1(a). On the other
hand $-1\leq K<0$ represents the situation when $p_{0}>p_{s}$, and the QP
density increases up to times $\theta\approx\tilde{\gamma}^{-1}$ - Fig. 1(b).
This situation is realized in MgB$_{2}$\cite{mgb2}.

\emph{(II) Superconducting state recovery dynamics.} At large times
($\theta\geq1/\tilde{\gamma}$) right hand side in Eqs.(\ref{RTdim1}%
,\ref{RTdim2}) is determined by the HFP decay term. To describe the recovery
we note that at large times the difference $q^{2}-p\approx\tilde{\gamma
}(p-p_{T})$ while $p(\theta)$ and $q(\theta)$ are slowly decreasing. We
introduce a function $s(\theta)=q(\theta)^{2}-p(\theta)$. Cleary $s(\theta)\ll
p,q$ and is slowly decaying ($ds/d\theta\ll s$). By neglecting the derivative
$ds/d\theta$ it follows from the definition of $s(\theta)$ that $2q\frac
{dq}{d\theta}=\frac{dp}{d\theta}$, which simplifies Eqs.(\ref{RTdim1}%
,\ref{RTdim2}) to%
\begin{equation}
(1+2\tilde{\gamma}+4q(\theta))\frac{dq(\theta)}{d\theta}=-2\tilde{\gamma
}(q(\theta)^{2}-q_{T}^{2})\text{ \ \ .}%
\end{equation}
After integration, we obtain an analytical solution:
\begin{equation}
-2\tilde{\gamma}\theta=(2+\frac{1+2\tilde{\gamma}}{2q_{T}})\ln{[\frac{q-q_{T}%
}{q_{s}-q_{T}}]}+(2-\frac{1+2\tilde{\gamma}}{2q_{T}})\ln{[\frac{q+q_{T}}%
{q_{s}+q_{T}}]} \label{qsFinal}%
\end{equation}
Combining solutions for prebottleneck and recovery dynamics we obtain an
approximate solution for $q(\theta)$ valid over the entire time range%
\begin{equation}
q(\theta)=q_{1}(\theta)+q_{2}(\theta)-q_{s}, \label{joint}%
\end{equation}
Here $q_{1}(\theta)$ and $q_{2}(\theta)$ are determined by Eqs.(\ref{PreQ})
and (\ref{qsFinal}) respectively. To illustrate that Eq.(\ref{joint})
describes the solutions of Eqs.(\ref{RTdim1},\ref{RTdim2}) we compare the
numerical solution with approximation given by Eq.(\ref{joint}). Fig.1 shows
excellent agreement for all the limiting cases.%

\begin{figure}
[h]
\begin{center}
\includegraphics[
trim=0.234943in 0.195299in 0.413464in 0.353893in,
height=7.8507cm,
width=8.1429cm
]%
{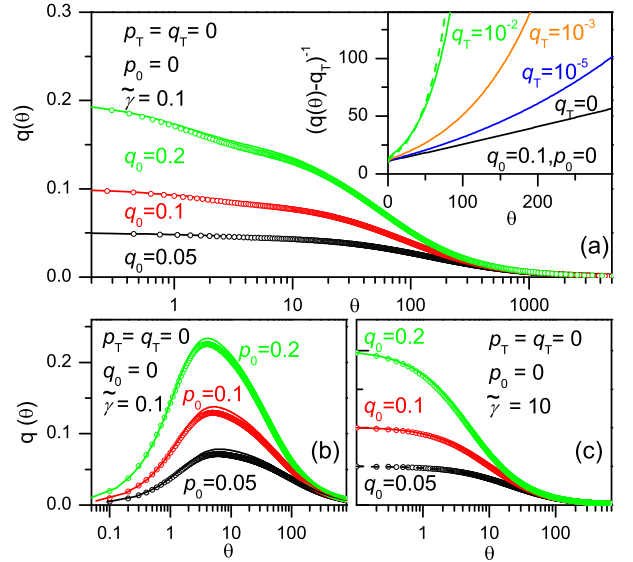}%
\caption{Analytical solutions of RT equations (open symbols) compared to
numerical solutions (solid lines) in several limiting cases at $T=0$ K. Panels
(a) and (b) represent the strong bottleneck limit ($\tilde{\gamma}=0.1$) for
different initial conditions (see text). Analytical solution for $q(\theta)$
is given by Eq.(\ref{joint}). Panel (c) presents the weak bottleneck case
($\tilde{\gamma}=10$). The inset to panel (a) presents the inverse of the
photoinduced QP density as a function of $\theta$ for various temperatures
($q_{T}$). At low-T ($q_{T}\longrightarrow0$) the time dependence is linear,
mimicking the bimolecular kinetics, while at higher T it becomes nearly
exponential (dashed curve presents an exponential decay fit). Indeed, similar
behavior was recently reported \cite{Kaindl}.}%
\end{center}
\end{figure}

Eq.(\ref{qsFinal}) can be simplified in the low and high-T limits:

\textit{i) Low temperature limit} $(q_{T}\ll1).$

In this case $2\ll\frac{1+2\tilde{\gamma}}{2q_{T}}$ and Eq.(\ref{qsFinal})
reduces to:
\begin{equation}
q(\theta)=q_{T}+\frac{2q_{T}(q_{s}-q_{T})\exp{(-\theta/\tau_{T})}}%
{(q_{s}+q_{T})-(q_{s}-q_{T})\exp{(-\theta/\tau_{T})}} \label{LTq}%
\end{equation}
The relaxation rate $\tau_{T}^{-{1}}=4\tilde{\gamma}q_{T}/(1+2\tilde{\gamma})$
is strongly reduced by the additional T-dependent factor $q_{T}\varpropto
n_{T}$. The behavior is consistent with the T-dependence of the relaxation
time $\tau$ in most of the SCs: cuprate\cite{mercury,Schneider,Segre} and
conventional \cite{mgb2,Tanner}. The low-T divergence of $\tau$ was originally
attributed to the bi-particle prebottleneck dynamics\cite{mercury} where
$\tau$ is divergent due to strong reduction of the thermal QP density. Here we
show that at low-T $\tau$ diverges also in the case of a well established bottleneck.

In the limit of $\theta/\tau_{T}\ll1$ Eq.(\ref{LTq}) further reduces to
$q(\theta)=q_{s}/(1+\frac{2q_{s}\tilde{\gamma}\theta}{1+2\tilde{\gamma}})$,
showing identical dynamics to that of the biparticle recombination - see inset
to Fig. 1(a). If one defines the relaxation rate as the slope $\tau
^{-1}=|dq(\theta)/d\theta|/(q_{s}-q_{T})$ at $\theta\rightarrow0$, as proposed
in Ref.\cite{orenstein}, the SC relaxation rate is intensity dependent
\begin{equation}
\tau^{-1}=\frac{2(q_{s}+q_{T})\tilde{\gamma}}{(1+2\tilde{\gamma})}\text{.}
\label{lowTtau}%
\end{equation}
The intensity dependence is observed only if $q_{s}\gg q_{T}$; in this case
$\tau^{-1}$ is proportional to the excitation intensity. It follows from
Eq.(\ref{lowTtau}), that the observation of pronounced intensity dependence of
$\tau$ is constrained only to the very low-T (determined by the $\Delta
/kT_{c}$ ratio) and is more likely to be observed in cuprates than in
conventional SCs. Indeed, the intensity dependent recovery dynamics has been
recently observed at low-T in BiSCO\cite{Kaindl}, in agreement with the
solution of RT equations in the strong bottleneck regime - inset to Fig. 1(a).

\textit{ii) High temperature limit} $(q_{T}\gtrsim1).$

In this case $2\gg\frac{1+2\tilde{\gamma}}{2q_{T}}$ and Eq.(\ref{qsFinal})
reduces to:
\begin{equation}
q(\theta)=\sqrt{q_{T}^{2}+(q_{s}^{2}-q_{T}^{2})\exp{(-\theta/\tau_{T})}}
\label{HTq}%
\end{equation}
Here the relaxation rate $\tau_{T}^{-1}=\tilde{\gamma}$ is weakly T-dependent
(due to intrinsic T-dependence of $\tilde{\gamma}$), and does not depend on
$q_{s}$ or $p_{s}$, implying that the relaxation rate is intensity
independent. The same is true also for $\tau^{-1}$.

\textbf{Weak bottleneck (}$\tilde{\gamma}\gg1$\textbf{).} In this case at
short times $\theta\approx\tilde{\gamma}^{-1}\ll1$ the last term in
Eq.(\ref{RTdim2}) is dominant, and the solution has the form
\begin{equation}
p(\theta)=p_{T}+(p_{0}-p_{T})\exp{(-\tilde{\gamma}\theta)}\text{ .}
\label{weakP}%
\end{equation}
The HFP density reaches its thermodynamic value on a time scale determined by
$\tilde{\gamma}^{-1}$. On such a short time scale QP density has not been
changed, therefore we can substitute $p$ with $p_{T}$ in Eq.(\ref{RTdim1}).
Taking into account that $q_{T}^{2}=p_{T}$ we obtain the solution:
\begin{equation}
q(\theta)=\frac{q_{T}[q_{0}+q_{T}+(q_{0}-q_{T})\exp{(-2q_{T}\theta)}]}%
{q_{0}+q_{T}-(q_{0}-q_{T})\exp{(-2q_{T}\theta)}}. \label{weakQ}%
\end{equation}
Eqs.(\ref{weakP},\ref{weakQ}) present the solution of Eqs.(\ref{RTdim1}%
,\ref{RTdim2}) for all timescales. As seen in Fig. 1(c) the difference between
the analytical and numerical solutions is negligible. Similar to the case of a
strong bottleneck the relaxation rate determined as the slope of the signal at
$\theta=0$ is found to be intensity dependent if $q_{0}\gg q_{T}$ with
$\tau^{-1}=(q_{0}+q_{T})$. The main difference between the strong and weak
bottleneck cases is that the absolute value of $\tau^{-1}$ is in the strong
bottleneck case reduced by the HFP decay time.

\textbf{Photoinduced QP density.\ }Earlier \cite{kabanov} we have discussed
the T and excitation intensity dependence of the photoinduced signal amplitude
\emph{Q} (proportional to the photoinduced QP density) assuming the bottleneck
condition and the energy conservation law. These results were recently
confirmed using a realistic phonon density of states\cite{nicol}. However,
solution of Eqs.(\ref{RTdim1},\ref{RTdim2}) allows unambiguous comparison of
\emph{Q}$(T)$ with the expected T-dependence in a SC - without an additional
assumption of the energy conservation as in Refs.\cite{kabanov,nicol}.

In the strong bottleneck regime initial dynamics leads to the stationary
densities of QPs and HFPs determined by Eqs.(\ref{Stat}). The initial
concentrations can be written as $q_{0}=q_{T}+\Delta q$ and $p_{0}%
=p_{T}+\Delta p$, where $\Delta q$, $\Delta p$ are photoinduced initial
concentrations of QPs and HFPs, respectively. Assuming weak perturbation limit
$\Delta p,$ $\Delta q\ll1$ the amplitude of the signal is given by:
\begin{equation}
\emph{Q}(T)\propto q_{s}-q_{T}\approx\frac{2\Delta p+\Delta q}{4\sqrt
{1+16p_{T}+8q_{T}}} \label{EqQ}%
\end{equation}
Taking into account that $p_{T}=q_{T}^{2}$ and $q_{T}=Rn_{T}/\eta$, and
normalizing $\emph{Q}$ to its low-T value, $A=\emph{Q}/\emph{Q}_{T\rightarrow
0K}$, we obtain $n_{T}\propto A^{-1}-1$. This relation is expected to be
general and valid for any SC irrespective on the gap symmetry allowing direct
estimation of the T-dependence of QP density in thermal equilibrium.

Importantly, in the weak bottleneck case $\emph{Q}$ should be proportional to
the pump intensity, and T-independent at low-T since $q(\theta=0)=q_{0}$,
while increasing when $T\rightarrow T_{c}$ due to closing of the SC gap. This
T-dependence has not been observed thus far.

In Fig. 2 we plot the T-dependences of $n_{T}$ (extracted from the raw data
via $n_{T}\propto A^{-1}-1$) and the SC recovery time $\tau$ for MgB$_{2}$ and
the cuprate superconductor Tl$_{2}$Ba$_{2}$CuO$_{6+\delta}$. The two datasets
capture the main features observed on other conventional \cite{Carr,Tanner} or
cuprate \cite{kabanov,Smith,mercury,THzAveritt,Segre,Schneider}
superconductors. In both \emph{Q} gradually decreases upon increasing T
towards T$_{c}$
\cite{kabanov,Smith,mercury,THzAveritt,Segre,Schneider,Carr,Tanner}, while the
low-T relaxation time increases when $T\longrightarrow0$. The only exception
thus far was the near optimally doped YBCO, where $\tau$ is roughly constant
at low-T \cite{kabanov,THzAveritt}. The main difference between cuprates and
conventional SC is a 1-2 orders of magnitude difference in $\tau$, which can
be understood in terms of an order of magnitude difference in the value of the
SC energy gap\cite{comparison}.

While at low-T and weak perturbations both strong and weak bottleneck
scenarios suggest slowing down of the relaxation upon cooling as well as
intensity dependent relaxation, the fact that the amplitude gradually
decreases upon T-increase suggests that all SC studied thus far are in the
strong bottleneck regime (in the weak bottleneck scenario \emph{Q} should
increase upon increasing T). In order to test the agreement of the model with
the data, we fit $n_{T}$ with Eq.(\ref{nTBCS}), where BCS form for $\Delta(T)$
was used. We find good agreement between the extracted values for
$2\Delta/T_{c}$ from the fit with the values from other experimental
techniques\cite{comparison,Tl2201gap}.%

\begin{figure}
[h]
\begin{center}
\includegraphics[
trim=0.215826in 0.235454in 0.216738in 0.393282in,
height=6.1132cm,
width=8.1495cm
]%
{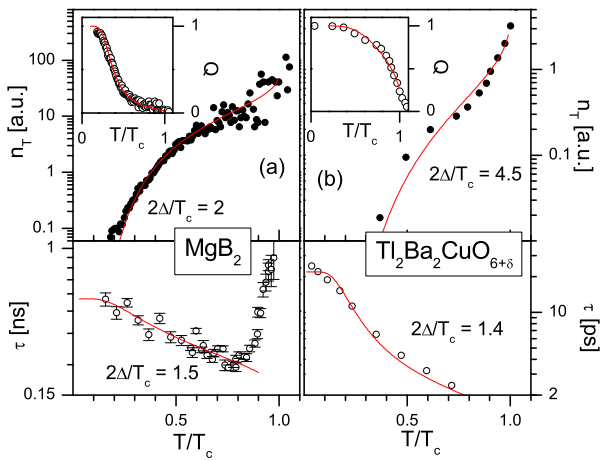}%
\caption{Comparison of the experimental data on T-dependence of $n_{T}$ and
$\tau$ on a) MgB$_{2}$ \cite{mgb2,CarrMgB2} and b) Tl$_{2}$Ba$_{2}%
$CuO$_{6+\delta}$ \cite{Smith} with the RT solution. $n_{T}$ is extracted from
the raw data (shown in insets) using $n_{T}\propto A^{-1}-1$ and fit with
Eq.(\ref{nTBCS}), while $\tau$ was fit with Eq.(\ref{lowTtau}). The extracted
values of $2\Delta/T_{c}$ ratio are also shown.}%
\end{center}
\end{figure}

We fit the T-dependence of the relaxation time $\tau$ with Eq.(\ref{lowTtau}).
Assuming $\tilde{\gamma}$ to be T-independent and expressing $q_{T}$ with
Eq.(\ref{nTBCS}), Eq.(\ref{lowTtau}) is rewritten to $\tau(T)=1/(\Phi+A(\Delta
T)^{1/2}\exp(-\Delta/T))$. Here $\Phi$ $=B(q_{s}-q_{T})$, while $A$ and $B$
are fitting parameters. Since the T-dependence of $q_{s}-q_{T}$ is measured
directly - see Eq.(\ref{EqQ}), we are left with 3 fitting parameters: $A$, $B$
and $\Delta$. The T-dependence of $\tau$ is governed mainly by $\exp
(-\Delta/T)$ term, while $A$ and $B$ account for the magnitude of $q_{T}$ and
$(q_{s}-q_{T})$ respectively. At the intermediate temperatures $q_{s}-q_{T}\ll
q_{T}$, and $\tau$ is governed by the T-dependence of $q_{T}$ showing
$\exp(\Delta/T)$ behavior. At low enough T, however, $q_{s}\gg q_{T}$ and the
relaxation time saturates. The values of $\Delta$ extracted from relaxation
time data are somewhat lower than the values extracted from the fit to $n_{T}%
$. This can be partially attributed to the fact that the T-dependence of
$\tilde{\gamma}$ was neglected. More importantly, improper (or missing)
treatment of the laser heating can also cause an underestimate of $\Delta$
obtained from the fit. Namely, due to laser excitation the equilibrium T of
the probed spot can be substantially higher than that of the cold
finger\cite{ACS}, especially at low-T. Therefore, the T-dependence of $\tau$
is not as steep as if the temperature scale was corrected for heating, giving
rise to an underestimate of $\Delta$. Note that heating is particularly
pronounced in experiments using high repetition rate laser systems like in
Ref.\cite{Smith}. Finally, we should note that the value of $\Delta$ extracted
is also somewhat dependent on the T-dependence of $\Delta$, i.e. when fitting
we used a BCS form. However, the accuracy of the available experimental data
is not sufficient enough to speculate about the deviation of the temperature
dependence from the BCS functional form.

In conclusion, we studied the evolution of a SC state following perturbation
with ultrashort optical pulses. We derived analytical solutions of RT model
for the strong and weak bottleneck regimes. Analytical solutions account for
both pair-breaking dynamics\cite{mgb2} as well as SC state recovery, and are
in excellent agreement with numerical results. They enable comparison to
experimentally measured quantities like the SC state recovery time and the
photoinduced QP density. Comparison with published data suggest that both
cuprate and conventional SCs follow the RT scenario, both being in the strong
bottleneck regime.

\textit{We would like to acknowledge A.S. Alexandrov and T. Mertelj for
critically reading the manuscript.}


\begin{thebibliography}{99}                                                                                               %


\bibitem {han}S.G. Han, et al., Phys. Rev. Lett. \textbf{65}, 2708 (1990).

\bibitem {Stevens}C.J. Stevens, et al., Phys. Rev. Lett. \textbf{78}, 2212 (1997).

\bibitem {DemsarYBCO}J. Demsar, et al., Phys. Rev. Lett. \textbf{82}, 4918 (1999).

\bibitem {kabanov}V.V. Kabanov, et al, Phys. Rev. B \textbf{59}, 1497 (1999).

\bibitem {Smith}D.C. Smith, et al., Physica C \textbf{341}, 2219 (2000).

\bibitem {THzAveritt}R.D. Averitt, \textit{et al.}, Phys. Rev.\textit{\ }%
\textbf{B 63}, 140502(R) (2001).

\bibitem {Segre}G.P. Segre, et al., Phys. Rev. Lett. \textbf{88}, 137001 (2002).

\bibitem {Schneider}M.L. Schneider, \textit{et al.}, Europhys. Lett.
\textbf{60}, 460 (2002).

\bibitem {mercury}J. Demsar, et al., Phys. Rev. B \textbf{63}, 054519 (2001).

\bibitem {Carr}G.L. Carr, \textit{et al.}, Phys. Rev. Lett. \textbf{85}, 3001 (2000).

\bibitem {Federici}J.F. Federici, \textit{et al.}, Phys. Rev. \textbf{B 46},
11153 (1992).

\bibitem {mgb2}J. Demsar, et al., Phys. Rev. Lett. \textbf{91}, 267002 (2003).

\bibitem {CarrMgB2}R.P.S.M. Lobo, \textit{et al.}, cond-mat/0404708.

\bibitem {Tanner}R.P.S.M. Lobo, \textit{et al.}, Phys. Rev. \textbf{B 72},
024510 (2005).

\bibitem {KaindlScience}R.A. Kaindl, et al., Science \textbf{287}, 470 (2000).

\bibitem {comparison}J. Demsar, et al., J. Supercond. \textbf{17}, 143 (2004).

\bibitem {orenstein}N. Gedik, et al., Phys. Rev. B \textbf{70}, 014504, (2004).

\bibitem {Kaindl}R.A. Kaindl, et al., CLEO/IQEC Technical Digest on CDROM,
(OSA, Washington, DC, 2004), ITuF1.

\bibitem {escape}A. Rothwarf, G.A. Sai-Halasz, and D.N. Langensberg, Phys.
Rev. Lett. \textbf{33}, 212, (1974).

\bibitem {roth}A. Rothwarf and B.N. Taylor, Phys. Rev. Lett. \textbf{19}, 27 (1967).

\bibitem {ovchkres}Yu.N. Ovchinnikov and V.Z. Kresin, Phys. Rev. B
\textbf{58}, 12416 (1998).

\bibitem {eliash}G.M. Eliashberg, Zh. Eksp. Theor. Fiz., \textbf{61}, 1254,
(1971) [Sov. Phys. JETP, \textbf{34}, 668 (1971)].

\bibitem {Maris}S. Tamura, H.J. Maris, Phys. Rev. B \textbf{51,} 2857 (1995).

\bibitem {Mihailovic}D.Mihailovic, et al., Phys. Rev. B \textbf{47}, 8910 (1993).

\bibitem {nicol}E. J. Nicol and J. P. Carbotte, Phys. Rev. B \textbf{67},
214506 (2003).

\bibitem {Tl2201gap}L. Ozyuzer et al., Phys. Rev. B \textbf{57,} R3245 (1998).

\bibitem {ACS}D. Mihailovic and J. Demsar in Spectrosopy of Superconducting
Materials, ed. Eric Falques, ACS Symposium Series 730; ACS, Washington, D.C., 1999.
\end{thebibliography}
\end{document}